\documentclass[pre,aps,10pt,twocolumn,showpacs]{revtex4}
\usepackage{graphics,graphicx,epsfig,color,subfigure,comment}

\begin{document}
\title{Universal fluctuations in KPZ growth on one-dimensional flat substrates}
\author{T. J. Oliveira}
\email{tiago@ufv.br}
\author{S. C. Ferreira}
\email{silviojr@ufv.br}
\author{S. G. Alves}
\email{sidiney@ufv.br}
\affiliation{Departamento de F\'isica, Universidade Federal de Vi\c cosa, 
36570-000, Vi\c cosa, MG, Brazil}

\date{\today}

\begin{abstract}

We present a numerical study of the evolution of height distributions (HDs)
obtained in interface growth models belonging to the Kardar-Parisi-Zhang (KPZ)
 universality class. The
growth is done on an initially flat substrate.  The HDs obtained for all
investigated models are very well fitted by the theoretically predicted Gaussian
Orthogonal Ensemble (GOE) distribution. The first cumulant has a shift that
vanishes as $t^{-1/3}$, while
the cumulants of order $2\le n\le 4$ converge to GOE as $t^{-2/3}$ or faster, behaviors previously observed in other KPZ systems.
These results yield a new evidence for the universality of the GOE distribution in
KPZ growth on flat substrates. Finally, we further show that the surfaces are
described by the Airy$_{1}$ process.
\end{abstract}
\pacs{68.43.Hn, 68.35.Fx, 81.15.Aa, 05.40.-a}

\maketitle


The last decade witnessed a great advance in the understanding of paradigmatic
nonequilibrium models for interface growth  by means of rigorous analytical
results~\cite{krugrev,SasaSpohnJsat} as well as their experimental realizations
~\cite{TakeSano,TakeuchiSP}.
Among the most prominent examples are several models belonging to
the Kardar-Parisi-Zhang (KPZ)
universality class introduced by the stochastic equation~\cite{KPZ}
\begin{equation}
 \frac{\partial h}{\partial t} = \nu \nabla^{2} h + \frac{\lambda}{2} (\nabla h)^{2} + \eta,
\label{eqKPZ}
\end{equation}
that describes the evolution of an interface $h(x,t)$. In this equation, $\eta$
is a white noise defined by $\langle \eta \rangle = 0$ and $\langle
\eta(x,t)\eta(x',t') \rangle = D \delta(x-x') \delta(t-t')$.  In $d=1+1$, the
KPZ equation yields self-affine surfaces with an interface width in a scale of
length $\ell$, defined as $w=\sqrt{\langle h ^2\rangle-\langle h \rangle^2}$,
given by the Family-Vicsek scaling ansatz~\cite{FV}: $w(\ell,t) =
t^\beta\mathcal{F}(\ell/t^{1/z})$, where $\mathcal{F}(x)\sim cont.$ for $x\gg 1$
and $\mathcal{F}(x)\sim x^{\beta z}$ for $x\ll1$. In $d=1+1$, the exponents of the KPZ class are
$\beta=1/3$ and $z=3/2$~\cite{KPZ}. Several models with the scaling exponents of
the KPZ universality class have been reported~\cite{barabasi,krug}.

For system belonging to the KPZ class in $d=1+1$, the Family-Vicsek scaling suggests an
interface height evolving as~\cite{krugrev}
\begin{equation}
 h(t) \simeq v_{\infty} t + sign(\lambda)(\Gamma t)^{1/3} \chi,
\label{eqht}
\end{equation}
where $v_{\infty}$ and $\Gamma$ are non-universal (model dependent) parameters,
and $\chi$ is a time independent random variable. The height distributions (HDs)
and consequently $\chi$  were computed exactly for some models in the KPZ class (see
Refs.~\cite{SasaSpohnJsat,krugrev} for recent reviews), strongly suggesting that
$\chi$ is a universal feature of the KPZ class. In a pioneer work,
Johansson~\cite{johansson} established a link between random matrix theory and
KPZ class by determining the exact asymptotic HDs of the totally asymmetric exclusion process (single step model) as the Tracy-Widom (TW) distribution of the
Gaussian unitary ensemble (GUE)~\cite{TW1}. In the same year, Pr\"ahofer and
Spohn~\cite{PraSpo1,PraSpo2} determined  the HDs of the radial polynuclear 
growth (PNG) model as a GUE distribution  whereas the Gaussian orthogonal
ensemble (GOE) distribution was obtained for the growth on a flat substrate.
Quite recently, exact solutions of the KPZ equation in $d=1+1$ corroborated GOE
distributions for flat~\cite{Calabrese} and GUE for radial~\cite{SasaSpo1,Amir}
geometries, respectively. 

Besides TW distributions for the heights, the limiting processes describing the
surfaces were identified as the so-called Airy processes \cite{krugrev}. For
circular growth, analytic solutions~\cite{PraSpo3}, experiments~\cite{TakeSano},
and simulations~\cite{SidTiaSil} strongly suggest that the limiting process
ruling the  KPZ universality class is given by the dynamics of the largest
eigenvalue in Dyson's Brownian motion of GUE matrices, the Airy$_2$ process.
Analogously, a limiting process for the flat case, named Airy$_{1}$, was found
for PNG~\cite{Boro1} and a similar model~\cite{Sasa1}, both belonging to the KPZ
class, indicating that it may be a universal feature of the KPZ class. It is
worth mentioning that Airy$_{1}$ process does not correspond to the dynamics of
largest eigenvalue in Dyson's Brownian motion of GOE matrices~\cite{Bornemann}.

KPZ universality class  in 1+1 dimensions was convincingly verified  in a
meticulous experiment involving two turbulent phases in the electroconvection of
nematic liquid crystal films~\cite{TakeSano,TakeuchiSP}. Using an unprecedented
statistics for experimental systems ($\sim10^3$ realizations), it was possible to observed not only the
KPZ scaling exponents $\beta=1/3$ and $z=3/2$, but also the GUE and GOE
distributions for radial and flat geometries, respectively. In both cases,
cumulants of order $n=2,3$ and 4 converge fast to the corresponding GOE and GUE
cumulants, but the mean ($n=1$) approaches the theoretical value as a power law
$t^{-1/3}$. This decay was also found in an analytical solution of the KPZ
equation in $d=1+1$ with an edge initial condition~\cite{SasaSpo1} and in
simulations of radial Eden model~\cite{SidTiaSil}. However, Ferrari and
Frings~\cite{Ferrari} have shown the non-universality of the amplitudes of this correction
and that higher order cumulants have no correction up to order
$\mathcal{O}(t^{-2/3}$).

The analysis of height distributions in computer simulations of
non-solvable models, supposedly belonging to the KPZ class, are relatively less
frequent than the analytical counterpart of solvable models. Indeed, some
reports for flat geometries dating from the beginning of nineties, when computer
resources were quite limited, demonstrate a surprisingly good agreement with the
lately found out theoretical GOE distribution~\cite{krug1,AmarPRA}. For a radial
geometry, we have recently observed asymptotic GUE distributions in computer
simulations of distinct Eden models grown from a single seed~\cite{SidTiaSil}.

In the present work, we revisit the computer simulations of non-solvable models
having  the KPZ exponents in $1+1$ dimensions. A flat initial condition is used.
We also integrated the KPZ equation numerically. In all cases, we confirmed that HDs
and cumulants agree with the GOE distribution. As recently observed for other
systems \cite{TakeSano,TakeuchiSP,SasaSpo1}, a correction in the mean
approaching zero as $t^{-1/3}$  was found while corrections in higher order
cumulants consistent with $t^{-2/3}$ were obtained for three of four
investigated models.  These results are in agreement with the corrections
obtained analytically for solvable KPZ models~\cite{Ferrari}. Finally, the
covariance of all investigated models agrees with the Airy$_{1}$ process, as
verified for some solvable models~\cite{Sasa1,Boro1}. 

We study the restricted solid-on-solid (RSOS) and  ballistic deposition (BD)
models~\cite{barabasi} as well as the numerical integration of Eq. (\ref{eqKPZ})
for two sets of parameters. In all cases, we use one-dimensional lattices of
size $L=2^{20}$ and periodic boundary conditions. Statistical averages were
performed over at least 400 independent samples. The RSOS model consists in the
deposition of a particle in a randomly chosen site. The deposition is rejected everytime the height
difference between nearest neighbors is greater than one lattice unity.
In the BD model, the deposition site is randomly chosen and the particle falls
normally to the substrate. When the first contact with the deposit happens, it
irreversibly sticks in this position.

\begin{table}[t]
\begin{center}
\begin{tabular}{cccccccccc}
\hline\hline
model    &  &  $v_{\infty}$   &  &  $\Gamma$      &  &    $A$     & \\
\hline
BD       &  &  $2.1398(7)$    &  &  $4.90(5)$     &  & $2.7(2)$  & \\
RSOS     &  &  $0.41903(3)$   &  &  $0.252(1)$    &  & $0.81(3)$  & \\
KPZ I    &  &  $0.05725(4)$   &  &  $0.00081(1)$  &  & $0.008(1)$     & \\
KPZ II   &  &  $0.08130(5)$   &  &  $0.00493(2)$  &  & $0.035(2)$     & \\
\hline\hline
\end{tabular}
\caption{Non-universal parameters for the investigated models.}
\label{table2}
\end{center}
\end{table}

We investigated a discrete one-dimensional KPZ equation given by~\cite{Moser}
\begin{eqnarray}
 h_j(t+\Delta t) &-& h_j(t) = \frac{\Delta t}{(\Delta x)^{2}} \left \lbrace \nu (h_{j-1} 
                             - 2 h_j + h_{j+1}) \right.  \nonumber \\
 &+& \left. \frac{\lambda}{8} (h_{j-1}-h_{j+1})^{2} \right \rbrace + \sigma \sqrt{12 \Delta t} R,
\label{eqKPZdisc}
\end{eqnarray}
where $\sigma\equiv\sqrt{2 D/\Delta x}$ and $R$ is a random number uniformly
distributed in the interval $[-0.5,0.5]$. Integration instabilities are
prevented by replacing the gradient term in KPZ equation by
$f(x)=(1-e^{-C|\nabla h|^{2}})/C$ where $C$ is a parameter to be adjusted to
control the instability~\cite{Dasgupta}. Integration was performed for fixed
parameters $\Delta x = 1$, $\Delta t = 0.04$ and $\nu = 0.5$. Two sets of
parameters were studied. In the first one, named as KPZ I, we used $\sigma=0.1$,
the coupling constant $g\equiv \lambda^{2} D/\nu^{3} =24$, and $C=0.5$. In the
second one, the KPZ II, we used $\sigma=0.2$, $g=12$ and $C=0.2$.

\begin{figure}[t]
\centering
\includegraphics[width=7.0cm]{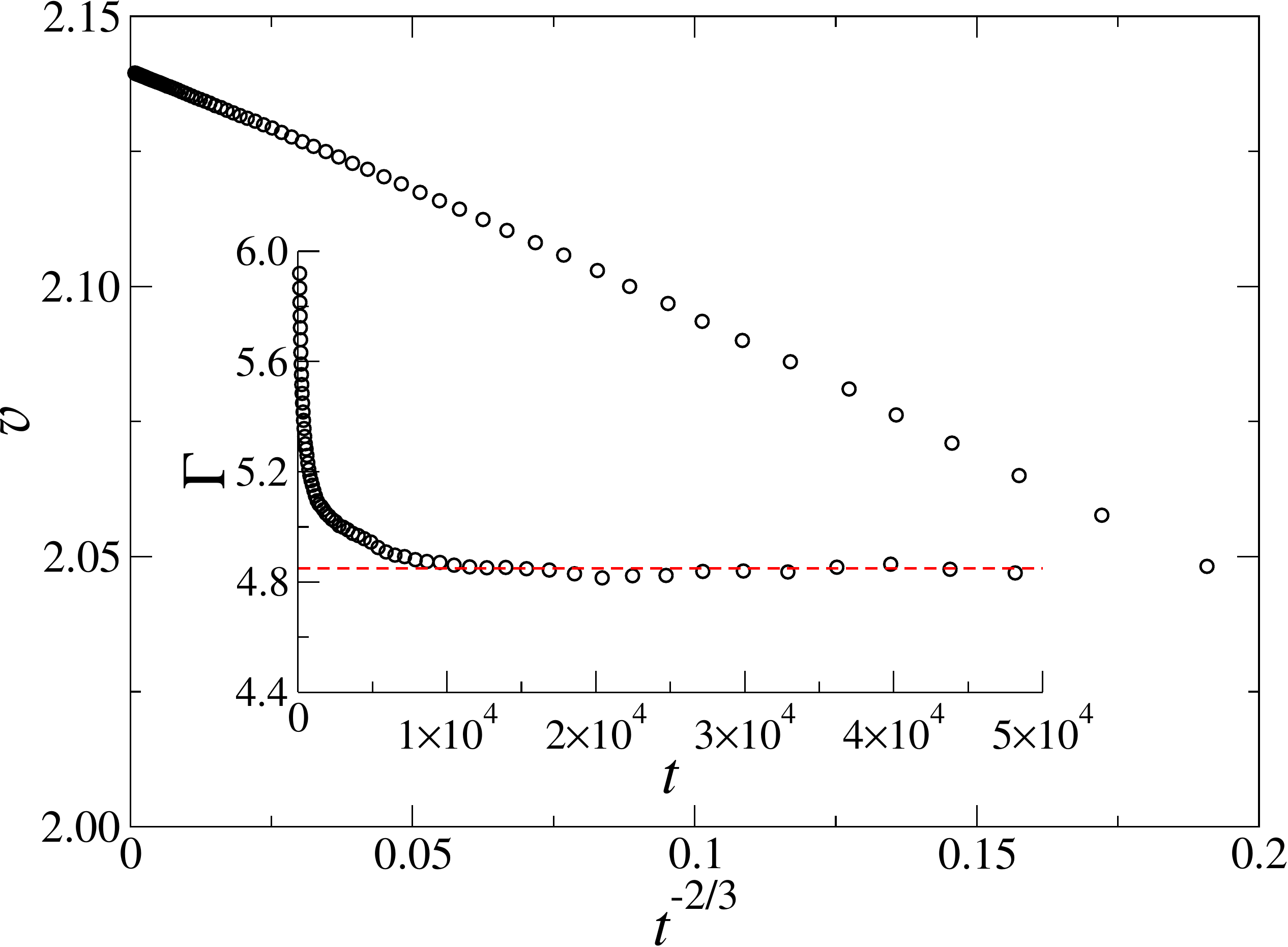}
\caption{\label{fig_nlp} (Color online) Growth velocity $v$ against $t^{-2/3}$
for the BD model. Inset shows $\Gamma$ against 
time for BD. Dashed line represents the estimate shown in Table~\ref{table2}.} 
\end{figure}

Comparison between HDs and random matrix distributions needs highly accurate
estimates of the parameters $v_{\infty}$ and $\Gamma$ given by Eq. (\ref{eqht}).
In BD and RSOS models studied here, estimates are reported in
literature~\cite{krug1}, but the accuracy is not sufficient for the aims of the
present work. Thus, we determined the parameters for all studied models. Using
Eq. (\ref{eqht}), the asymptotic growth velocity can be obtained from the
extrapolation of plots $v\equiv d \langle h \rangle/ d t$ against $t^{-2/3}$
when $t\rightarrow\infty$, as illustrated in Fig.~\ref{fig_nlp} for the BD
model. The asymptotic velocities for all investigated models are presented in
Table~\ref{table2}. Notice that our estimates for RSOS and BD models are
consistent with those formerly reported in Ref.~\cite{krug1}: $v_{\infty} =
0.419$ and $2.14$. The parameter $\Gamma$ was obtained from the second order
cumulant of the HDs. Accordingly Eq. (\ref{eqht}), it reads as $\langle h^{2}
\rangle_{c} \simeq (\Gamma t)^{2/3} \langle \chi^{2} \rangle_{c}$, where
$\langle x^n\rangle_c$ is the $n$th cumulant of $x$. Since we expect that $\chi$
fluctuations are given by the GOE distribution, we used $\langle \chi^{2}
\rangle_{c} = \langle \chi_{GOE}^{2} \rangle_{c}=0.63805$~\cite{PraSpo1} in our
calculations. The insertion to Fig. \ref{fig_nlp} shows $\Gamma \simeq
(\langle h^{2} \rangle_{c}/\langle \chi_{GOE}^{2} \rangle_{c})^{3/2}/t$ as a
function of time for the BD model. The estimated $\Gamma$ values for all investigated
models are presented in Table \ref{table2}. Plots similar to Fig.~\ref{fig_nlp}
were obtained for all models.

\begin{figure}[t]
\includegraphics[width=7.0cm]{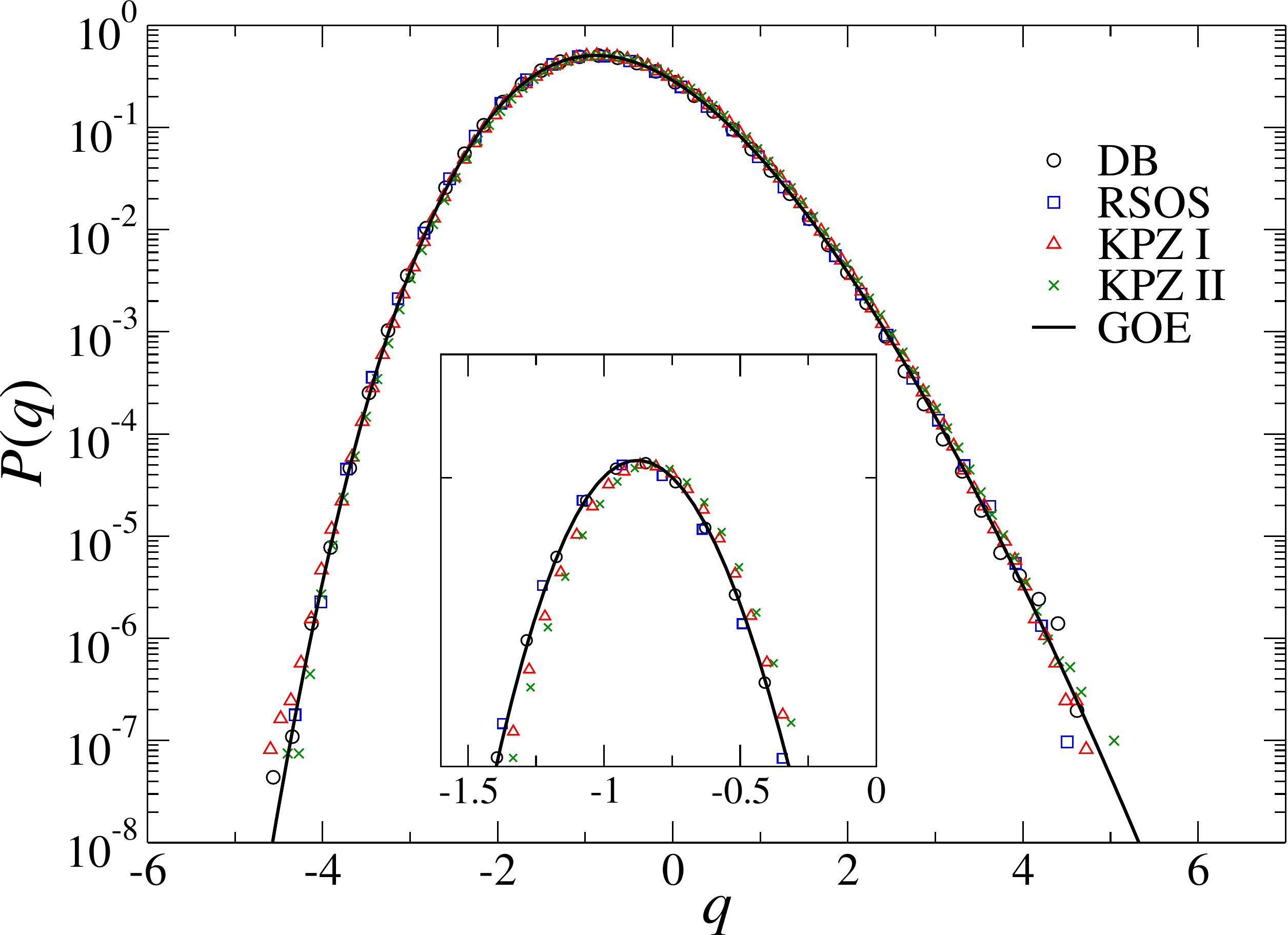}
\caption{(Color online) Height distributions scaled accordingly Eq.~(\ref{eqht})
for all investigated models. Simulation time of $t=2\times 10^4$ was used for BD
and RSOS and $t=800$ for the integrated KPZ equation. The solid line is the GOE
distribution. The inset shows a zoom around the peak of the distributions.}
\label{fig_diskpz}
\end{figure}

\begin{figure}[t]
\includegraphics*[width=8.0cm]{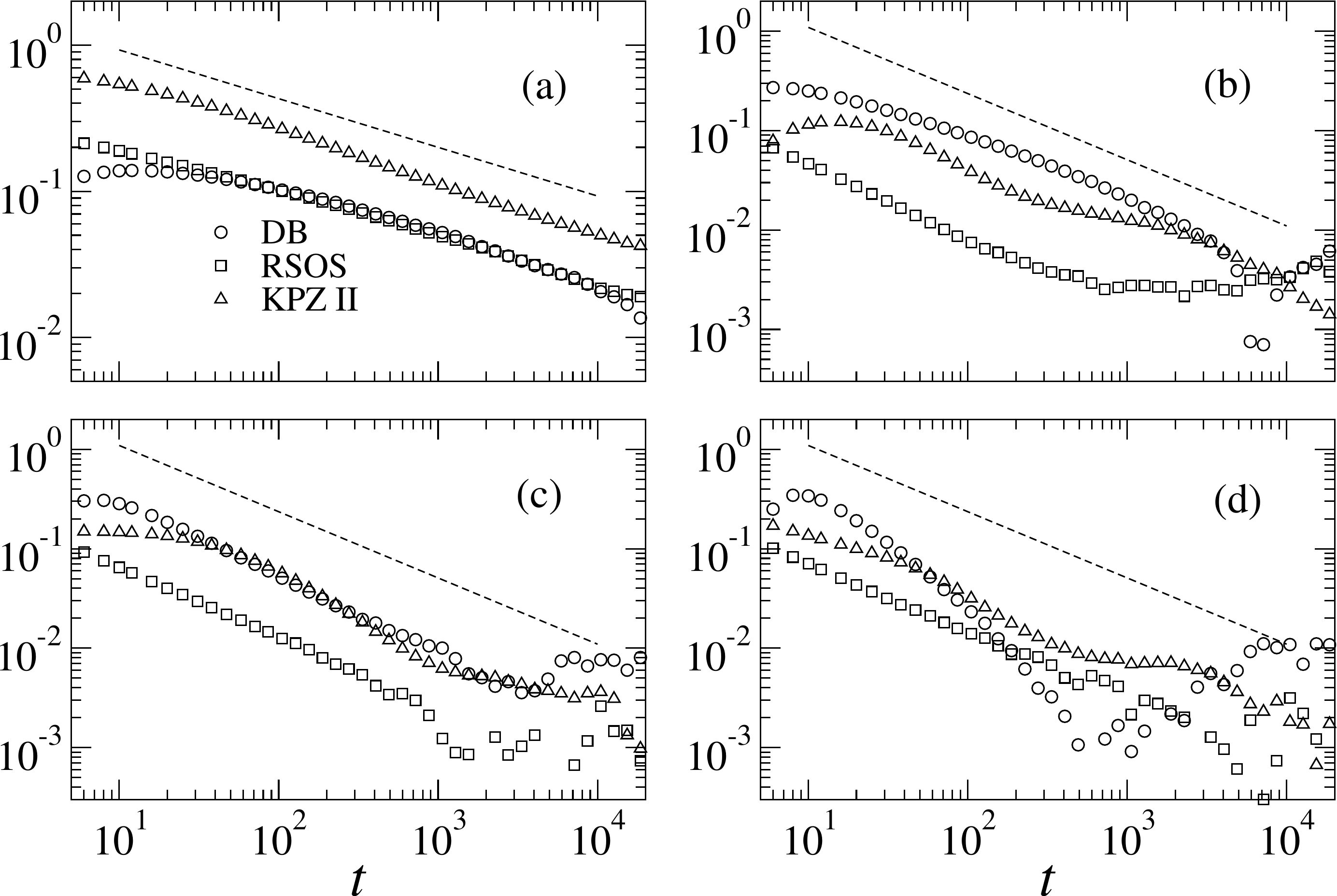} \\~\\
\includegraphics*[width=6.0cm]{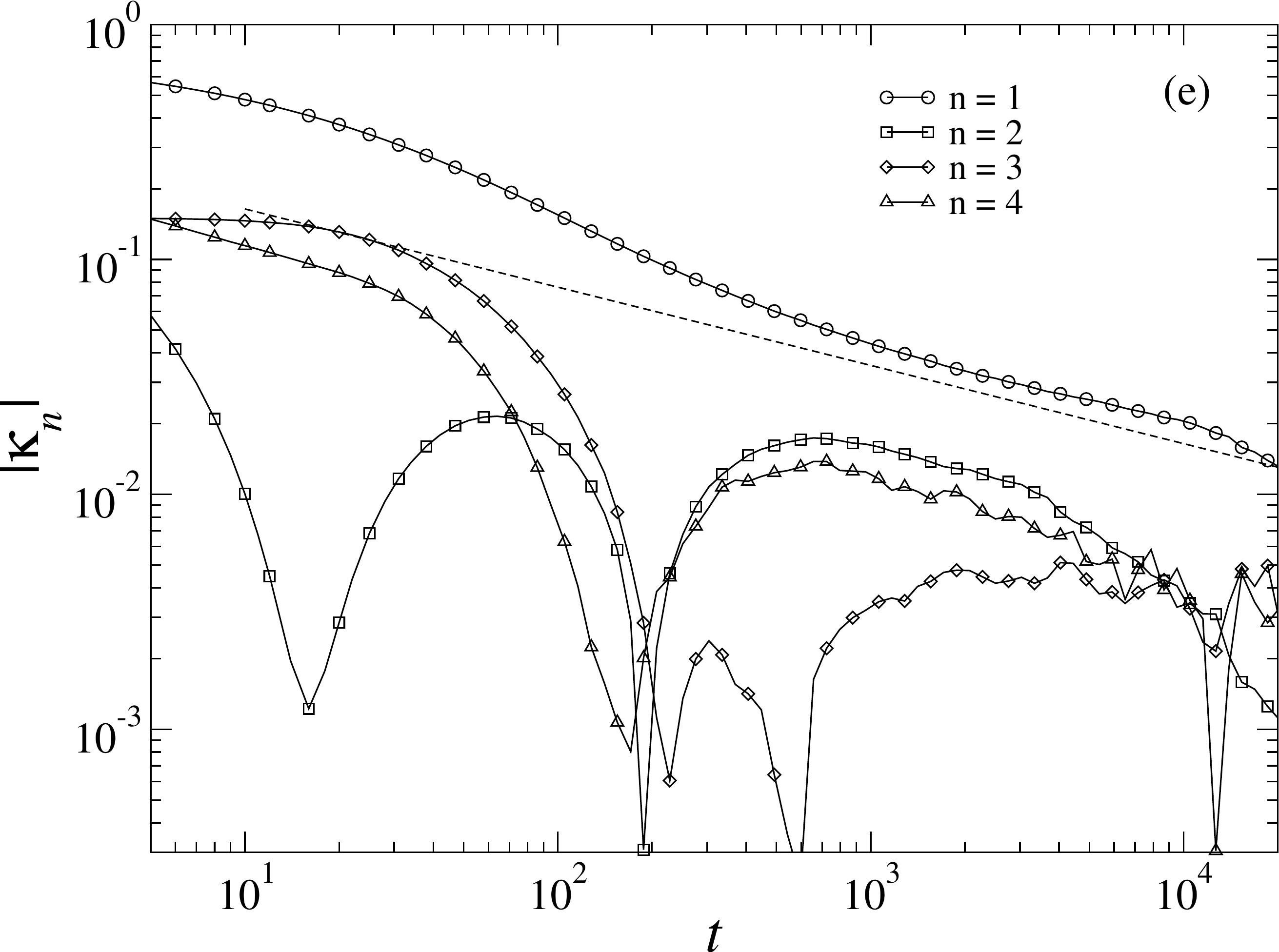}
\caption{Absolute differences between cumulants of the scaled HDs and GOE
distribution against time. Results for (a) first, (b) second, (c) third, and (d)
fourth order cumulants are shown. Results for KPZ I are shown apart in (e). The
dashed lines have slopes -1/3 in (a) and (e) and -2/3 in (b)-(d).} 
\label{fig_cumu}
\end{figure}

We study the rescaled HDs following the KPZ ansatz given by Eq.~(\ref{eqht}). If
this equation holds, the fluctuations of $q \equiv (h - v_{\infty} t)/(\Gamma
t)^{1/3}$ would be given by the GOE distribution. Fig. \ref{fig_diskpz} shows
the rescaled distributions $P(q)$ for all investigated models. As expected, the
agreement with GOE is noticeable, but the distributions are slightly shifted in
relation to GOE, as highlighted in the inset of Fig. \ref{fig_diskpz}.

\begin{figure}[bt]
\centering
\includegraphics[width=7.0cm]{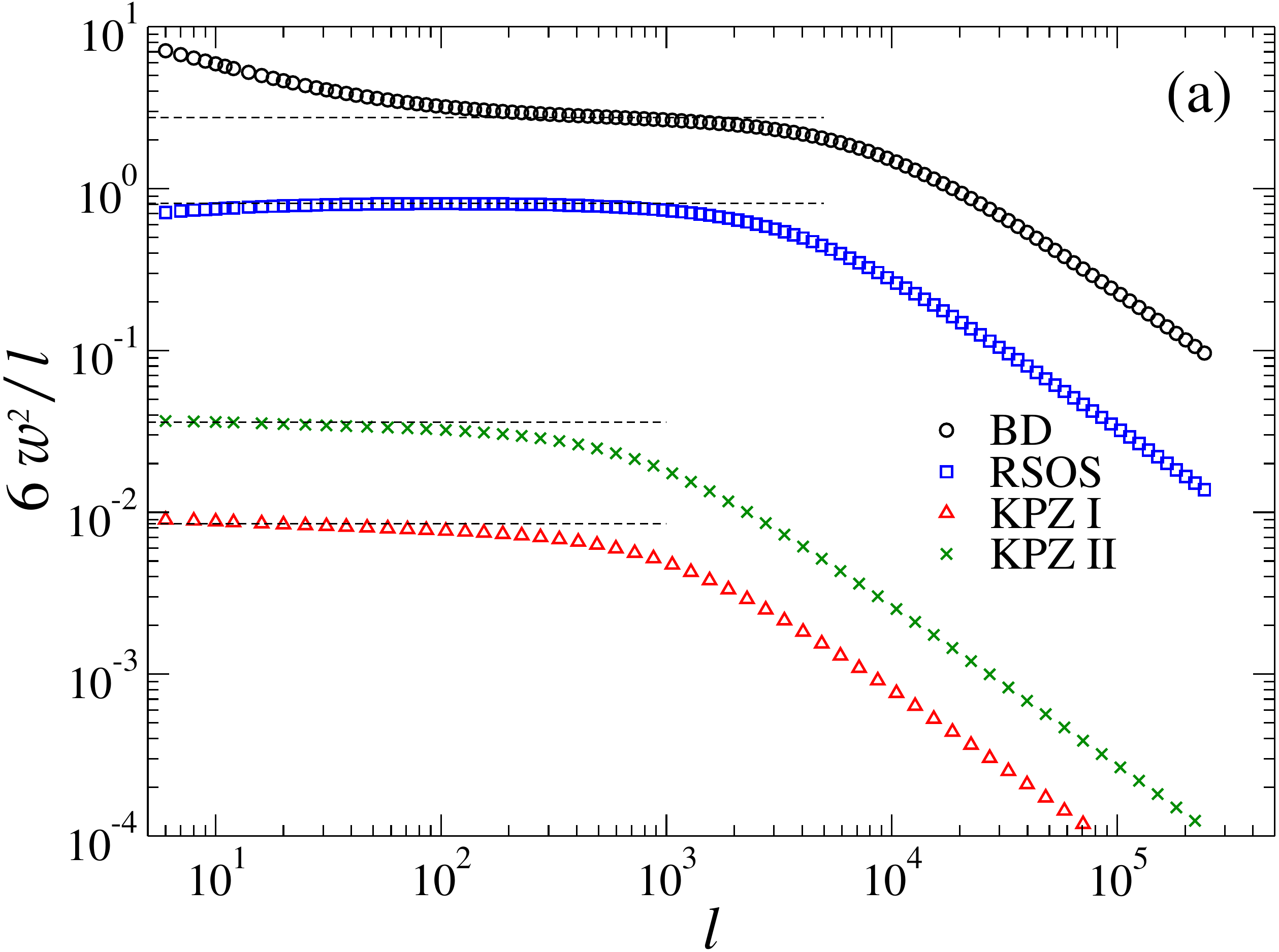} \\~~\\
\includegraphics[width=7.0cm]{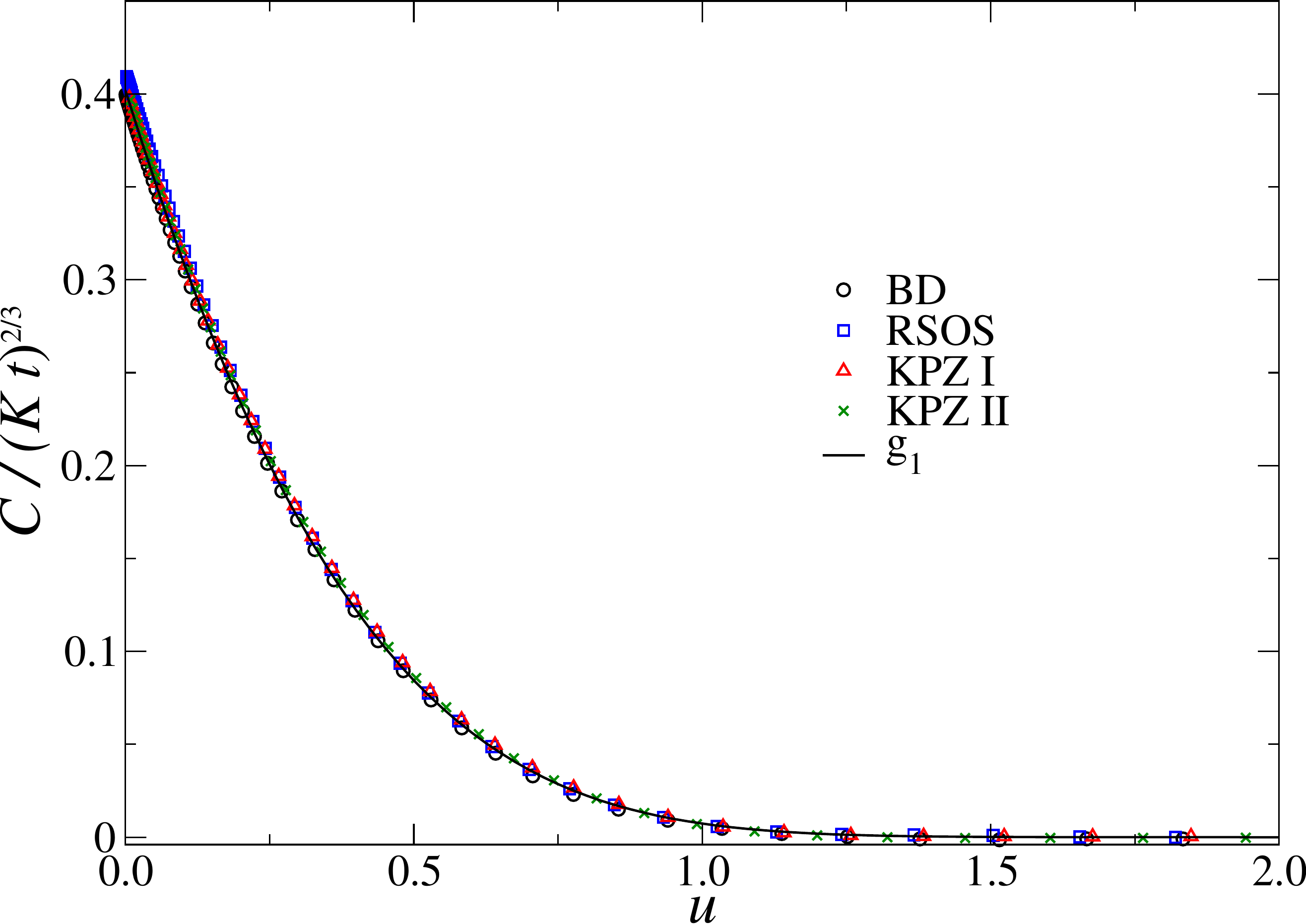}
\caption{(Color online) (a) Rescaled local roughness for different models.
Simulation times are $t=5\times 10^{4}$ for BD and RSOS and $t=2\times 10^3$ for
KPZ I and II. (b) Scaled two-point correlation function against $u \equiv (A
l/2)/(K t)^{2/3}$ for different models. Time is $t=10^4$ for BD and
RSOS and $t=400$ for the integrated KPZ equation. The solid line is the
covariance of the Airy$_1$ process.}
\label{fig_covar}
\end{figure}

The absolute differences between $P(q)$ and GOE cumulants, $|\kappa_n|$, are
shown as functions of time in Fig.~\ref{fig_cumu}. The first cumulant difference
decays closely to a power law $t^{-1/3}$ for all investigated models, except for KPZ
I. In the latter, the initial decay is faster but seems to converge to
$t^{-1/3}$. Ferrari \textit{et al}.~\cite{Ferrari} have shown that the corrections in the first
cumulant for a number of solvable models belonging to the KPZ class read as
\begin{equation}
\kappa_1 = \langle q\rangle - \langle \chi \rangle = at^{-1/3}+\mathcal{O}(t^{-2/3}) 
\label{eq:kappa1}
\end{equation}
where $a$ is a model dependent constant that can even be null for some
models. Numerically, if the factor $a$ is much smaller than the factor of higher
order correction, one must observe a fast initial decay~[$\mathcal{O}(t^{-2/3})$]
followed by a $t^{-1/3}$ decay. This  is  observed in the KPZ I results shown in
Fig.~\ref{fig_cumu}(e). For KPZ I, the higher order cumulants reach GOE very
fast and then the differences $|\kappa_n|$ fluctuate around zero. Furthermore, the
non-universality of the constant $a$ is verified in our simulation since $a$ is
negative for BD and RSOS and positive for the KPZ equation integration. Ferrari
\textit{et al}~\cite{Ferrari} have also shown that the higher order cumulants have no
correction up to order $\mathcal{O}(t^{-2/3})$. Our simulations are in agreement
with this claim as one can see in Figs. \ref{fig_cumu}(b)-\ref{fig_cumu}(e), where some curves
are consistent with a $t^{-2/3}$ decay while others decay faster, which is
particularly evident in KPZ I case. The conclusion of Ref.~\cite{Ferrari}
stating the non-universality of the correction refers only to the amplitudes
since the scenario suggested by Eq.~(\ref{eq:kappa1}) is quite general: any
correction to the first moment, if it exists, should decay with an exponent
$-1/3$, while corrections in higher order cumulants, if they exist, must decay as
fast as or faster than $t^{-2/3}$.  

Besides the universality of the height fluctuations, the limiting process
describing the interface is an important issue on surface growth. To
check if our simulations yield surfaces given by an Airy$_1$ processes, we studied
the two-point correlation function 
\begin{equation}
 C(l,t) \equiv \left\langle h(x+l,t) h(x,t)  \right\rangle 
- \left\langle h  \right\rangle^{2},
\end{equation}
that is expected to scale as $C(l,t) \simeq (K t)^{2/3}
g_{1}(u)$ with $u \equiv (A l/2)/(K t)^{2/3}$, where
$g_{1}(u)$ is the covariance of the Airy$_{1}$ process and $K = \lambda A^{2} =
2\Gamma$~\cite{PraSpo3}. 
The amplitude $A$ can be obtained from the squared local roughness
\begin{equation}
 w^{2}(l,t) = \langle h^{2}(x,t) \rangle_{l}  - 
\langle h(x,t) \rangle_{l}^{2} \simeq A l/6,
\end{equation}
where $\langle \cdots \rangle_{l}$ means averages over windows of size
$l$~\cite{krug1,TakeSano}. Fig. \ref{fig_covar}(a) shows the plots
used to determine the amplitudes shown in Table~\ref{table2}. The short plateau for the BD model is due to strong finite time effects of this model.
Notice that the amplitude is given by $A = D/\nu$ for the KPZ equation, implying
that $A=0.01$ and $A=0.04$ are expected for KPZ I and II, respectively. The
slightly different amplitudes observed are due to the approximated integration
method~\cite{LamShin}. 
Figure \ref{fig_covar}(b) shows the results for the rescaled two-point
correlation function exhibiting a very satisfactory agreement among all models
and the covariance of the Airy$_{1}$ process ($g_{1}$). This confirms that all
models are described by the Airy$_1$ process and gives one more evidence of this
universal feature in KPZ  universality class. Similar results were obtained
at different times, indicating small finite-time effects in the two-point
correlation function.

In conclusion, we investigated non-solvable models in the KPZ universality
class in one-dimensional flat substrates as well as a numerical integration of
the KPZ equation. We analyzed the height distributions and the two-point
correlation function. Height fluctuations of all investigated models are
described by GOE distributions as conjectured for the KPZ class for flat
substrates, and previously observed in analytical~\cite{Calabrese} and
experimental~\cite{TakeuchiSP} works.  The cumulant analysis yields a correction
vanishing as $t^{-1/3}$ for the first cumulant whereas the corrections for
higher order cumulants go to zero as $t^{-2/3}$ or faster. The latter result
is in agreement with  theoretical analysis of Ferrari \textit{et al}~\cite{Ferrari}. Finally,
we have found that the limiting processes describing the surface is the Airy$_1$
process in agreement with previous results for solvable models~\cite{Boro1,Sasa1}.

\begin{acknowledgments}
This work was partially supported by the Brazilian agencies CNPq, FAPEMIG, and
CAPES.  Authors thank F. Bornemann by kindly providing the covariance of the
Airy$_1$ process.
\end{acknowledgments}



\end{document}